\documentclass[twocolumn,nofootinbib,preprintnumbers,floatfix,aps,showpacs,prl]{revtex4}

\usepackage[hyperfootnotes=false,bookmarks=false]{hyperref}

\usepackage{graphicx}
\usepackage{amsmath}
\usepackage{multirow}
\usepackage{float}
\usepackage{color}

\newcommand{\be}{\begin{equation}}
\newcommand{\ee}{\end{equation}}
\newcommand{\bea}{\begin{eqnarray}}
\newcommand{\eea}{\end{eqnarray}}

\begin{document}

\title{Peak locations and relative phase of different decay modes of the $a_1$ \\axial vector 
resonance in diffractive production}

\author{Jean-Louis Basdevant}
\email{{jean-louis.basdevant@polytechnique.edu}}
\affiliation{Physics Department, Ecole Polytechnique, 91128 Palaiseau, France}
\author{Edmond L. Berger}
\email{berger@anl.gov}
\affiliation{High Energy Physics Division, Argonne National Laboratory, Argonne, Illinois 60439, USA}

\begin{abstract}
We show that a single $I=1$ spin-parity $J^{PC}= 1^{++}$ $a_1$ 
resonance can manifest itself as two separated mass peaks, one decaying into an S-wave $\rho\pi$ system and the second 
decaying into a P-wave $f_0(980)\pi$ system, with a rapid increase of the phase difference between their amplitudes arising mainly 
from the structure of the diffractive production process.  This study clarifies questions related to the mass, width, and decay rates of the 
$a_1$ resonance raised by the recent high statistics data of the COMPASS collaboration on $a_1$  production in 
$\pi N \rightarrow \pi \pi \pi N$ at high energies.  
\end{abstract}


\maketitle

New insight into the properties of light mesons is emerging from the unprecedented statistical precision of the COMPASS experiment at CERN  where beams of 190 GeV pions interact with nucleon targets~\cite{Adolph:2015pws}.  These data are bound to enrich (if not challenge) 
our understanding of low-energy meson spectroscopy while, in addition, uncovering possible evidence for long-sought states 
of the strong interaction QCD potential beyond the quark-antiquark states of the standard model. 

We focus on the isospin $1$ axial-vector resonance $a_1$ (reported in the Particle Data Compilation as $a(1)(1260)$~\cite{Agashe:2014kda}).  Evidence is presented in the COMPASS data for a new narrow $J^{PC} = 1^{++}$ axial-vector state, strongly coupled to the $\pi f_0(980)$ system.  This observation of a  peak in the two body $\pi f_0$ P-wave intensity at a mass of 1.42 GeV, combined with a phase motion close to $180^\circ$ with respect to other waves, appears at face-value to mean that a second axial-vector resonance is present, close in mass to the known broad $a_1(1260)$ that couples mainly to the $\pi \rho$ system~\cite{Adolph:2015pws}.  While these three features, i.e. two peaks at different masses and a rapid phase variation, are clearly  present, there are reasons to be surprised, among which we mention: 
(i) The $a_1(1260)$ is a central member of the axial-vector nonet, which, together with the $J^{P}= 0^{-},\, 1^{-},\,\textrm{and}\;  0^{+}$ form the ground-state of the light quark-antiquark spectrum.  A newcomer in the family would be difficult to accommodate. 
(ii) It is peculiar to have two $J^{PC} = 1^{++}$ three-pion states, with identical quantum numbers, close in mass (within a full width of each other), with orthogonal decay modes, without the presence of some new quantum number.  The $K^0_S-K^0_L$ system led to decisive discoveries in fundamental physics; neutrino mixing is a spectacular current example.  However, in the $a_1$ case, we see no candidate for a distinguishing quantum number.   

Our basic approach to high energy forward production of three pion states in pion-nucleon interactions is the Drell-Hiida-Deck mechanism~\cite{Drell:1961zza}.  This model has been studied extensively in the production of the $J^P=1^+$ $\rho\pi$ system~\cite{Berger:1976nr}, and here we extend the analysis to the $J^P=1^+$ $f_0(980)\pi$  system.  An important difference is that whereas the $\rho\pi$ system is in an orbital S-wave state, the $f_0(980)\pi$ is in an orbital P-wave state.  Since the two-body $\rho \pi$ and $f_0 \pi$ systems are strongly interacting, we must modify the Deck mechanism with the proper final-state interactions due to the re-scattering  of these systems.  This is an inescapable physical consistency condition of the entire analysis. 
The unitary coupled channel approach that we developed in Refs.~\cite{Basdevant:1977ya, Basdevant:1978ty, Basdevant:1978tx}, should be an ideal way to show whether one resonance suffices or 
whether the COMPASS data do require two nearby resonances with the same axial-vector quantum numbers in the three-pion system.  

In this Letter, we demonstrate that a \emph{single} resonance suffices to explain the data and that the $f_0 \pi$ decay mode of the usual $a_1$ is  being observed for what appears to be the first time~\cite{Agashe:2014kda}.  Our method can be used to determine new values for the mass and width of the $a_1$, information important for lattice QCD and other calculations of the hadron spectrum.   

\begin{figure}[h]
  \begin{center}
  \includegraphics[width=0.5\textwidth]{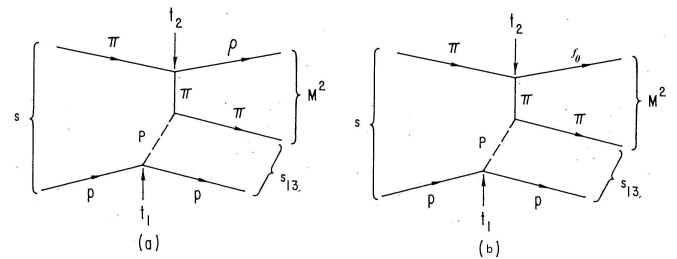}
  \caption{Deck production processes for (a) $\rho\pi$ and (b) $f_0\pi$. }
\label{fig:deckf}
\end{center}
\end{figure}
{\em Two-channel Deck amplitudes.}
We follow closely Refs.~\cite{Berger:1976nr} and~\cite{Basdevant:1977ya}.
We consider the Deck amplitudes for production of the quasi-two body systems $\pi \rho$ and $\pi f_0$ at small 
momentum transfer and high incident energy.  We denote these 
$T^{\rho}_D = T_D(\pi \textrm{N} \to \pi \rho \textrm{N})$ and 
$T^f_D = T_D(\pi \textrm{N} \to \pi \textrm{f}_0\, \textrm{N})$. 
The reactions are represented in Figs.\ref{fig:deckf} (a) and (b). 

The $\pi \rho$ case has been studied at length. Its amplitude (c.f., Eq.~(2.1) of Ref.~\cite{Basdevant:1977ya}) is 
\begin{equation}
T^{\rho}_D= g_{\rho\pi\pi}K_\rho(t_2)\frac{1}{m^2_\pi -t_2} {i}s_{13}e^{b t_1}\sigma_{\pi p}
\label{DCK}
\end{equation}
where $g_{\rho\pi\pi}$ is the $\rho\pi\pi$ coupling constant ($g^2/(4\pi) =2.4$), $K_\rho$ is the magnitude of the incident pion momentum in the 
$\rho$ rest frame~\cite{Berger:1976nr}, $b$ is the slope of the $\pi N$ elastic 
diffraction peak, and $\sigma_{\pi p}$ is the $\pi p$ total cross-section. The invariants
$s_{13}$, $t_1$ and $t_2$ are labeled in Fig.~\ref{fig:deckf}~(a).

Similarly, the $\pi f_0$ production amplitude is 
\begin{equation}
T^{f}_D= g_{f_0\pi\pi}\frac{1}{m^2_\pi -t_2} {i}s_{13}e^{b t_1}\sigma_{\pi p}\quad.
\label{FCK}
\end{equation}
Choosing the average value of the 
$f_0 \rightarrow \pi\pi$ width of $60$~MeV, we obtain a numerical value $g_{f_0\pi\pi} \simeq 1.45$~GeV. The other factors 
in Eq.~(\ref{FCK}), relative to  Fig.~\ref{fig:deckf}~(b) have the same meaning as in Eq.~(\ref{DCK}). 

The $\rho \pi$  Deck background has been well studied, where, by background, we mean the amplitude before any final state interactions are included.  We refer to Ref.~\cite{Basdevant:1977ya} and extract what is useful in the present analysis. 
We work in the final $\rho \pi$ ($f_0 \pi$) center of mass frame; $M$ is the invariant mass of this system. 
In the limit of forward production $(t_1\to 0)$ and large $s$, Eq. (\ref{DCK}) produces the $\rho \pi$ system predominately in an 
S-wave~\cite{Berger:1976nr},   
used in previous calculations (e.g., Ref.~\cite{Basdevant:1977ya}).   
  
However, the $J^P=1^+$ $f_0\pi$ system is in an orbital P-wave.  To address $f_0\pi$, we must extend the partial wave extraction calculations to 
finite values of $t_1$ and $s$.  We present the complete calculation of these amplitudes elsewhere~\cite{Basdevant:2015ysa}.  The important feature is that the higher partial wave amplitudes are of order $t_1/M^2$ or $M^2/s$ with respect to the dominant S-wave.  
An immediate consequence is that $f_0\pi$ P-wave production should have a noticeably smaller rate than the $\rho\pi$ S-wave process, 
as is borne out in the complete calculation and exhibited by the COMPASS data, where the intensity of the $f_0\pi$ peak at 1.42\,GeV, is lower than that of the  $\rho\pi$ peak at 1.26\,GeV by a factor of the order of a few $10^{-3}$.

In the COMPASS experiment, the value of the square of the invariant total energy is 
$s= 380$~GeV$^2$ while the momentum-transfer $t_1$ in the smallest bin is $t_1 \in [-0.1,-0.13]$~GeV$^2$.    
We are interested in values of $M\sim 1 $ to $2$~GeV.  Since $\vert t_1 \vert/ M^2 \gg M^2/s$, the only relevant kinematic corrections come from the momentum transfer dependence.  We choose to work at the fixed value $t_1=-0.1$~GeV$^2$, and we checked that within the first t-bin ($t_1 \in [-0.1,-0.13]$~GeV$^2$), our results do not vary appreciably.
A convenient dimensionless expansion parameter is
\begin{equation}
\Theta_1 = \frac{t_1}{(M^2-m_{\pi}^2)} .
\label{tetard}
\end{equation}

The $J^P=1^+$ S-wave $\rho\pi$ background amplitude is, to first order in $\Theta_1$,
\bea
T^{Deck}_S&=& -\frac{s}{(M^2-m^2_\pi)}\times \nonumber\\ 
&&\left( 1-\frac{1}{2}{\Theta_1}(\frac{(3M^2+m^2_\pi)}{(M^2-m_\pi^2)}- \frac{E_{\rho}}{E_\pi})(\frac{1}{y}\ln\frac{1+y}{1-y})
\right)
\quad,
\label{ropib}
\eea
where $E_\pi$ and $E_\rho$ are the pion and  $\rho$ energies in the $\rho\pi$ rest frame, and where 
$y= p_\pi/E_\pi$ 
is the $\rho\pi$ phase space factor, $p_\pi$ being the pion momentum in the $\rho\pi$ rest frame.

The  $J^P=1^+$ P-wave $f_0\pi$ amplitude is, at the same order in $\Theta_1$,
\bea
 T^{Deck}_P&=&+\frac{3}{2}\frac{s}{(M^2-m^2_\pi)} {\Theta_1}\times \nonumber\\
&&\left(\frac{(3M^2+m^2_\pi)}{(M^2-m^2_\pi)}- \frac{E_{f_0}}{E_\pi}\right) (\frac{-2}{y} +\frac{1}{y^2}\ln(\frac{1+y}{1-y}))
\quad, \label{fop}
\eea
where $E_\pi$, $E_{f_0}$ are the pion and $f_0$ energies, $p_{\pi}$ the pion momentum in the $f_0\pi$ rest frame and, as above, $y=p_{\pi}/E_\pi$.

Equation~(\ref{fop}) is a major clue to our investigation.  The right hand side contains the factor 
${(3M^2+m^2_\pi)}/{(M^2-m^2_\pi)}- {E_{f_0}}/{E_\pi}$.  This factor is negative at low values of $M$ (since $m_{f_0}/m_\pi > 3)$, 
but it vanishes near $M\simeq 1.38$~GeV and becomes positive afterward.  Furthermore, if we give this term some small imaginary part, 
its phase will switch suddenly from $-180^\circ$ to zero.  This sudden and rapid phase variation is not a dynamical effect in the sense of a resonant phase, but it originates in the structure of the dynamical process by which the $f_0\pi$ state is produced.
Another interesting qualitative feature of Eq.~(\ref{fop}) is that it grows in the region of interest ($M\sim 1.2 $ to $1.4$~GeV) and therefore tends to push a resonance peak upward in $M$.  

Keeping in mind the parameters introduced in Eqs.~(\ref{DCK}) and (\ref{FCK}), our two $J^{PC}=1^{++}$  amplitudes are 
\begin{equation}
\left( \begin{array}{l}
T_{Deck} (\rho\pi)\\
T_{Deck} (f_0\pi)
\end{array}\right) = \frac{2i\sqrt{2}s N}{(M^2-m_\pi^2)}\left( \begin{array}{l}
g_{\rho\pi\pi} K_\rho\sigma_{\pi p} \tilde T_{\rho\pi}\\
g_{f_0\pi\pi}\;\sigma_{\pi p}\; \tilde T_{f_0\pi}\end{array}\right) ,
\label{tdkrf}
\end{equation}
where, $\tilde T_{\rho\pi}$ and $\tilde T_{f_0\pi}$ can be read off from Eqs.~(\ref{ropib}) and (\ref{fop}).   The structure remains the same after 
we unitarize.  The normalization factor $N$ is irrelevant for present purposes and is taken equal to $1$ here.\\  

{\em Unitarization.}
For theoretical and technical details about multi-channel final state unitarization, we refer to the literature, in particular to Ref.~\cite{Babelon:1976kv} where the general analysis may be found and to Ref.~\cite{Basdevant:1977ya} where a specific application is made.
We recall that if $\mathcal {S}$ is the (two-channel) strong interaction $S$ matrix, then the unitarized  Deck amplitude $\mathcal {T_D}$, which we can write as a two dimensional vector as in Eq.(\ref{tdkrf}), has a right hand unitarity cut along which it satisfies the relation 
\begin{equation}
\mathcal{T_D}^+ = \mathcal S \mathcal{T_D}^-
\label{unita}
\end{equation}
$\mathcal{T_D}^+$ and $\mathcal{T_D}^-$ being the values of the unitarized Deck amplitude above and below the cut.

Our basic assumption is that there is \emph{a single $a_1$ resonance whose (unique) second-sheet pole parameters we determine}.
Since we are dealing with a two-channel case, we parametrize the coupled $\rho\pi$ and $f_0\pi$ final state interactions (or rescattering) via this resonance.  In order to do this, we introduce a $K$ matrix, as in Eq.(3.14) of Ref.~\cite{Basdevant:1977ya}:  
\begin{equation}
 K(M^2)=\left( \begin{array}{cc}
\frac{g_1^2}{s_1-M^2}&\frac{g_1g_2}{s_1-M^2}\\
\frac{g_1g_2}{s_1-M^2}&\frac{g_2^2}{s_1-M^2}\\
\end{array} \right) \quad.
\label{Kmat}
\end{equation}

The crucial tool to treat coupled channel final state interactions is a $D$-matrix, related directly to the $S$ matrix.  It is presented explicitly in Eq.(3.15) of~\cite{Basdevant:1977ya}: 
\begin{equation}
D(M^2)= \frac{1}{\mathcal{D}_0(M^2)}\left( \begin{array}{cc}
g_1 & -g_2(s_1-M^2-\alpha^2C_2)\\
g_2 & g_1(s_1 -M^2- \alpha^2 C_1) \end{array}\right)
\label{Dmat}
\end{equation}
where $\alpha^2= g_1^2+g_2^2$, $C_1$ and $C_2$ are Chew-Mandelstam functions~\cite{Chew:1960iv}, and the energy denominator function $\mathcal D_0(M^2)$ is 
\begin{equation}
\mathcal D_0(M^2)= (s_1-M^2- g_1^2 C_1(M^2)-g_2^2 C_2(M^2))\quad.
\label{Dfunc}
\end{equation}
The function $\mathcal D_0(M^2)$ contains all the information (that we put in) on the coupled-channel $\rho\pi$-$f_0\pi$ strong interaction.  It is 
an analytic function which possesses the $\rho\pi$ and $f_0\pi$ branch cuts from $[m_\rho+m_\pi]^2$ to infinity and from $[m_{f_0}+m_\pi]^2$ to infinity. Its second sheet  pole determines the nominal position and width of the $a_1$ resonance.  

As in \cite{Basdevant:1977ya}, \emph{the unitarized Deck amplitude with resonant rescattering corrections taken into account} is 
\bea
T_D^u(M^2)&=& T_D(M^2) - \frac{1}{\pi}D(M^2)\times \nonumber\\
&&\int_{(m_\rho+m_\pi)^2}^\infty\, ds'\,\frac{ImD(s') T_D(s')}{(s'-M^2)}\quad.
\label{tdu}
\eea
Here $T_D^u(M^2)$ is a two-dimensional vector and $T_D(M^2)$ is the ``background" Deck amplitude discussed above.\\

{\em Direct production contribution.}
In addition to its manifestation through final state interactions, the $a_1$ may also be produced \emph{directly} in a diffractive process $\pi p \rightarrow a_1 p$.  For direct production, we choose 
\begin{equation}
T_{dir}(s,M^2) = \frac{is \sigma_{\pi p}G}{\mathcal D_0(m^2)}\left(\begin{array}{l}
f_1 \\ f_2\end{array} \right) ,
\label{dirprod}
\end{equation}
where $G$ represents the diffractive coupling to the $\pi$ to the $a_1$, and $f_1, f_2$ are the couplings of the $a_1$ in the $\rho\pi$ and $f_0 \pi$ 
channels respectively.   Our final amplitude is
\begin{equation}
T(M^2)= T_D^u(M^2) + T_{dir}(s,M^2)\quad .
\label{tdd}
\end{equation}
In the one-channel case exemplified by $\rho$ photo-production~\cite{Soding:1965nh}, the interference of these two terms can shift the apparent peak 
position of the $\rho$.  

{\em Analysis and results.}
Some salient points can be made short of a detailed fit to data.   
We first select appropriate values of the $a_1$ parameters that provide a good global description. 
The COMPASS results fix these parameters more stringently than when we dealt only with the S-wave $\rho\pi$ system (and other S-wave channels, such as $K^* \bar K$).  Here, the acceptable mass and width of the $a_1$, \emph{defined by the position of the second sheet pole}, turn out to be quite restricted.  Our analysis indicates that:
\begin{center}
$M(a_1) \simeq 1.40\pm 0.02~{\rm {GeV}}$, \\

$\Gamma(a_1) \simeq 0.30\pm 0.05~\rm {GeV}$. 
\end{center}
These values of mass and width correspond to values of the  parameters $s_1 \sim  2.002$~GeV$^2$ and $g_1 \sim   0.732$~GeV.  

The  interesting parameter to vary is the ratio $\gamma=g_2/g_1$ in order to find the range of values that produce two peaks with appropriate characteristics:  the $f_0\pi$ peak occurs at higher mass than the $\rho\pi$ peak, and the ratio of maximal intensities of these peaks, i.e. 
$\rho\pi$/$f_0\pi$, falls between 1,000 and 500, as indicated by the available data. 
These requirements lead to \emph{negative} values of $\gamma$ in the range $[-0.1, -0.055]$. In other words, the $a_1$ couplings to $\rho\pi$ and to $f_0\pi$ have opposite signs.  We choose as our central value $\gamma=g_2/g_1= -0.08$.   
\begin{figure}[h]
\begin{center}
{\includegraphics[width=.45\textwidth]{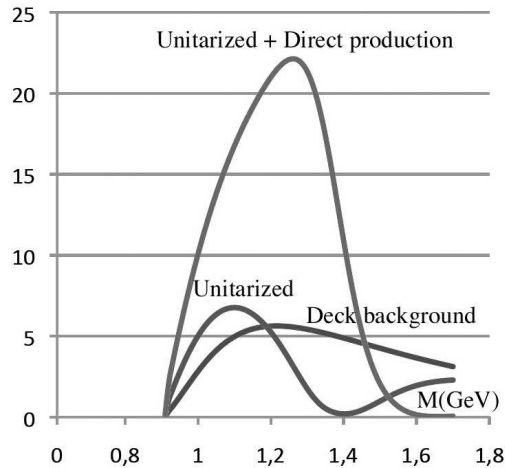}}
\caption{The $\rho\pi$ differential cross section as a function of the mass $M$, in three cases: background Deck, unitarized Deck as in Eq.(\ref{tdu}), and the final result including direct production. }
\label{rhovs}
\end{center}
\end{figure}

The next step is to determine the amount of direct production necessary to fix the two peaks, one in $\rho\pi$, the other in $f_0\pi$, at their desired positions, i.e. $M=1.26$~GeV for  $\rho\pi$ and $M=1.42$~GeV for $f_0\pi$. 
Values such as $G\sigma_{\pi p} f_1 = 120$ and $G\sigma_{\pi p} f_2= 5.5$  
in Eq.~(\ref{dirprod}) ensure good positions for the two peaks, and this situation is stable when one varies the parameter $\gamma$.
The ratio of direct production to the background Deck amplitude is consistent with the value we obtained previously~\cite{Basdevant:1977ya} in our analysis of data at much lower energies, with smaller statistics.  

\begin{figure}[h]
\begin{center}
{\includegraphics[width=.5\textwidth]{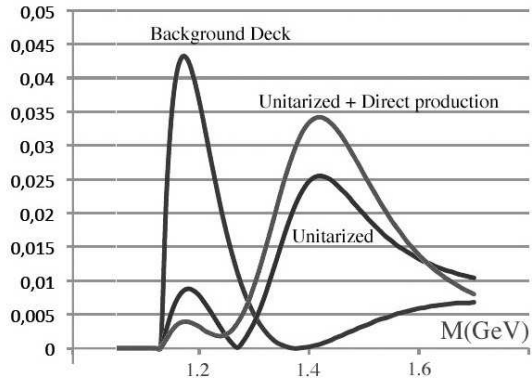}}
\caption{The $f_0\pi$ differential cross section as a function of the mass $M$, in three cases: background Deck, unitarized Deck as in Eq.(\ref{tdu}), and the final result including direct production. }
\label{f0vs}
\end{center}
\end{figure}
It is interesting to see how the two peaks are built up.  
We plot in Fig.~(\ref{rhovs}) the shapes of the $\rho \pi$ intensity as a function of the energy $M$, for the various terms in the calculation.
The pure Deck background does not produce a resonant shape. The unitarized amplitude shows effectively the $\sim \cos\delta$ zero that appears in the one-channel case~\cite{Basdevant:1977ya} (to which this problem is actually very close). Finally, direct production produces the observed peak, at the right  position.  A similar set of curves for the  $f_0\pi$ intensity is shown in Fig.~(\ref{f0vs}). Notice that the form of the Deck background by itself appears to simulate a narrow resonance peak at threshold (of course without any accompanying phase).
\begin{figure}[h]
\begin{center}
{\includegraphics[width=.45\textwidth]{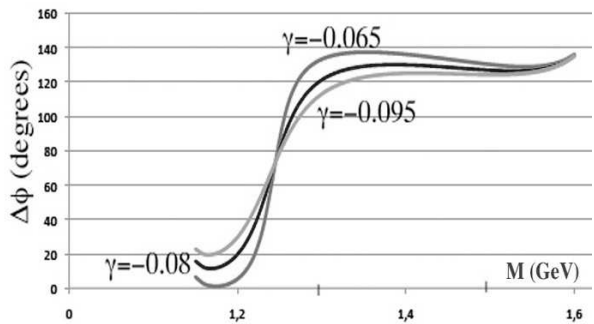}}
\caption{Three phase-differences between $f_0\pi$ and $\rho\pi$, in the 1.2 to 1.6 GeV region. The middle curve corresponds to our ``central" solution $\gamma=-0.08$; the two others (marked) to other values in the range of interest. In all three cases, the other results of the calculation remain practically unchanged: peak positions and intensities.}
\label{phas}
\end{center}
\end{figure}

The separation of the positions of the two peaks is evident.  The width of $a_1(1260)$ is about twice the width of $a_1(1420)$ in the calculation.  The $a_1(1420)$ peak is also more symmetrical, with width about $0.14$~GeV.  The lower end of the $f_0\pi$ intensity exhibits a (tiny) peak at around $1.2$~GeV owing to the zero emphasized in Eq.~(\ref{fop}).  

We display in Fig.~(\ref{phas}) a set of phase differences between the $f_0\pi$
and $\rho\pi$ amplitudes for three values of our parameter $\gamma$.  We adopt 
$\gamma=-0.08$ as our ``central" value (subject to more refined analyses).
We call attention to the rapid rise of over $100^\circ$  in the phase difference in the mass 
region $1.2$ to $1.3$~GeV.   
A quantitative fit to the data would yield a precise value of $\gamma$ and of the branching ratio of the $a_1$ into 
$f_0\pi$ of the order of $10^{-3}$ relative to the dominant decay mode $a_1\rightarrow \rho\pi$.  

{\em Summary.}  We find that the main features of the COMPASS data, two mass peaks separated by $\sim 160$~MeV with significant relative phase motion, are fully compatible with a \emph{single} $a_1$ resonance.  A detailed quantitative fit of the data with this formalism would lead to a new  determination of the mass and width of the $a_1$ and of its branching fraction into $f_0 \pi$.  

{{\em Acknowledgments.}
We thank  Stephan Paul and other members of the COMPASS collaboration for bringing this interesting topic to our attention. 
JLB thanks Khosrow Chadan, Philippe Boucaud and Jean-Pierre Leroy, for their considerable help at the Laboratoire de Physique Th\'eorique d'Orsay.  ELB is supported in part by the U.S. DOE under Contract No. DE-AC02-06CH11357.}

\end{document}